\documentclass[article, twocolumn,
   aps, pra,
  amsmath,amssymb,
  longbibliography,
  ]{revtex4-1}
\usepackage{graphicx,color}
\usepackage{amsmath}
\usepackage{natbib}
\usepackage{epsfig}
\begin{document}

\title{Modified Bridgman formula for the thermal conductivity of complex (dusty) plasma fluids }

\author{S. A. Khrapak}\email{Sergey.Khrapak@gmx.de}
\affiliation{Joint Institute for High Temperatures, Russian Academy of Sciences, 125412 Moscow, Russia}
\author{A. G. Khrapak}
\affiliation{Joint Institute for High Temperatures, Russian Academy of Sciences, 125412 Moscow, Russia}

\begin{abstract}
A simple and popular Bridgman's formula predicts a linear correlation between the thermal conductivity coefficient and the sound velocity of dense liquids. Unfortunately, it cannot be applied to strongly coupled plasma-related fluids, because the sound velocity can greatly increase as screening weakens. We propose a modification of the Bridgman formula by correlating the thermal conductivity coefficient with the transverse (shear) sound velocity. This approach is demonstrated to work reasonably well in screened Coulomb (Yukawa) fluids and can be useful in the context of complex (dusty) plasmas.        
\end{abstract}

\date{\today}

\maketitle

About a century ago Bridgman proposed a simple formula relating the coefficient of thermal conductivity with the sound velocity and density of dense liquids~\cite{Bridgman1923},
\begin{equation}\label{tc1}
\Lambda = \frac{2c_s}{\Delta^{2}},
\end{equation}
where $\Lambda$ is the coefficient of thermal conductivity, $c_s$ is the sound velocity, $\Delta=\rho^{-1/3}$ is the mean separation between the molecules, and $\rho$ is the liquid number density. In our consideration temperature is measured in energy units, so that from Fourier's law for the heat flow, $q=\Lambda (dT/dx)$, it is clear that $\Lambda$ is measured in cm$^{-1}$s$^{-1}$. The Boltzmann constant is thus set unity $k_{\rm B}T\rightarrow T$. The rationale behind Bridgman's formula is that the energy between neighbouring molecules is transferred at the speed of sound and each molecule controls the characteristic area of $\Delta^2$. The numerical coefficient in Eq.~(\ref{tc1}) is disputable and the values between $2$ and $3$ have been used in the literature~\cite{BirdBook,XiCPL2020,ZhaoJAP2021,KhrapakPRE01_2021}. 

Bridgman's expression was apparently the first theoretical formula applied to explain the thermal conductivity of dense fluids, but it still remains a useful simple reference correlation often discussed in contemporary literature~\cite{ZhaoJAP2021,XiCPL2020,KhrapakPRE01_2021,Chen2021}. One of the present authors has recently performed a systematic analysis of correlations between the thermal conductivity coefficients and sound velocities in simple model fluids (hard-sphere and Lennard-Jones), monatomic liquids (argon and krypton), diatomic liquids (nitrogen and oxygen), and several polyatomic liquids (water, carbon dioxide, methane, and ethane)~\cite{KhrapakJMolLiq2023}. It has been demonstrated that linear correlations are well reproduced for model fluids as well as real monatomic and diatomic liquids. The correlations seem less convincing in polyatomic molecular liquids. The coefficient of proportionality in Eq.~(\ref{tc1}) is not fixed: it is about unity for monatomic liquids and generally increases with molecular complexity.  Even if not truly universal, Bridgman's formula remains quite appealing, because the information about the sound velocity in various liquids is either easily accessible or can be relatively easily measured experimentally.

However, there is a serious deficiency in the Bridgman's approach. It cannot be applied to plasma-related fluids with soft and long-range interactions between the particles. To be concrete, let us concentrate on complex (dusty) plasmas -- systems of charged particles immersed in a neutralizing plasma medium~\cite{TsytovichUFN1997,FortovBook,FortovUFN,FortovPR,BeckersPoP2023}. The electron contribution to the thermal conductivity, which can certainly dominate because of high electron mobility in many realistic situations, is not addressed here. Instead, we focus on how the energy is transported through the particle component. This mimics heat transfer  process in ordinary single-component fluids with soft pairwise interactions. The advance of complex plasma in this context is that heat transfer can be relatively easy measured in experiment, by analyzing video-recorded particle dynamics~\cite{NunomuraPRL2005_HT,FortovPRE2007,NosenkoPRL2008,NosenkoPoP2021}.  

In the first approximation, the particles in complex plasmas are interacting with the screened Coulomb (Debye-H\"uckel or Yukawa) potential of the form
\begin{equation}
\phi(r)=\frac{Q^2}{r}\exp\left(-\frac{r}{\lambda}\right),
\end{equation}    
where $Q$ is the particle charge, $\lambda$ is the screening length (equal to the plasma Debye radius in the simplest case), and $r$ is the distance between a pair of particles. The properties of Yukawa systems are  described by the two dimensionless parameters: the Coulomb coupling parameter $\Gamma=Q^2/aT$ and the screening parameter $\kappa=a/\lambda$, where $a=(4\pi\rho/3)^{-1/3}$ is the Wigner-Seitz radius. The screening parameter $\kappa$ determines the softness of the Yukawa potential. The limit $\kappa\rightarrow 0$ corresponds to the extremely soft and long-ranged Coulomb potential $\propto 1/r$, operating in the one-component plasma (OCP)~\cite{BrushJCP1966,BausPR1980}.
In the context of complex plasmas the relatively ``soft'' regime with $\kappa\sim {\mathcal O}(1)$ is usually realized. 

To explain why the original Bridgman's formula cannot be applied to plasma-related systems, let us consider the weakly screened regime $\kappa\lesssim 1$. In this regime the sound velocity is virtually independent of the coupling strength and tends to the conventional dust acoustic wave (DAW) velocity~\cite{KhrapakPRE03_2015,KhrapakPPCF2015,KhrapakPoP10_2019}. Using the definition of the DAW velocity~\cite{Rao1990} we get
\begin{equation}
c_{\rm s}\simeq c_{\rm DAW}=\omega_{\rm p}\lambda\propto \kappa^{-1},
\end{equation}
where $\omega_{\rm p}=\sqrt{4\pi Q^2\rho/m}$ is the plasma frequency and $m$ is the particle mass. Although the dust acoustic velocity remains finite for any specified set of dusty plasma parameters, it can increase indefinitely as $\kappa$ decreases. At the same time, the range of wave-vectors to which acoustic dispersion applies becomes very narrow for small $\kappa$, see e.g. Fig.~1 from Ref.~\cite{OhtaPRL2000} as an illustration. This is quite natural, because in the OCP limit the longitudinal (compressional) collective excitations exhibit a plasmon dispersion, rather than a conventional acoustic one in the long-wavelength limit. The sound velocity is thus infinite in this limiting OCP case. Since the thermal conductivity coefficient of the OCP fluid remains finite~\cite{ScheinerPRE2019}, we conclude that Eq.~(\ref{tc1}) is irrelevant in this case.   

The purpose of this Letter is to modify Eq.~(\ref{tc1}) in such a way that it becomes applicable to plasma-related systems with soft pairwise interactions. The idea is very simple, we propose to substitute the conventional sound velocity in Eq.~(\ref{tc1}) by the {\it transverse sound velocity} $c_t$. Transverse (shear) modes are supported in dense liquids, although with a so-called ``$k$-gap'' --  a forbidden long wavelength region for $k<k_{\rm gap}$, where the real frequency is zero~\cite{OhtaPRL2000,PramanikPRL2002,NunomuraPRL2005,BandyopadhuayPLA2008,HosokawaPRL2009,HosokawaJPCM2015,GoreePRE2012,BolmatovSciRep2016,
BrykJCP2017,YangPRL2017,KryuchkovSciRep2019,KhrapakJCP2019,HosokawaPRL2009}. The transverse sound velocity is related to the infinite-frequency (instantaneous) shear modulus via $G_{\infty}=m\rho c_t^2$. The instantaneous shear modulus can be expressed as integral containing the first two derivatives of the pairwise interaction potential and the radial distribution function (RDF) $g(r)$~\cite{ZwanzigJCP1965}. For Yukawa systems several simple approaches to estimate $G_{\infty}$ across coupling regimes have been proposed~\cite{KhrapakIEEE2018,KhrapakPoP10_2019,KhrapakPoP02_2020}. 

This idea may seem absurd at first glance, but it is not for several reasons. The transverse velocity does not exhibit divergence for soft interaction potentials, as $c_{\rm s}$ does. 
The reduced transverse sound velocity increases monotonously with the coupling parameter and reaches a quasi-universal value at the fluid-solid phase transition~\cite{KhrapakPRR2020}. 
So does the reduced thermal conductivity coefficient of simple fluids in the high density regime~\cite{KhrapakPoF2022,KhrapakJCP2022_1}.

The transverse sound velocity has been recently identified as an isomorph invariant of Yukawa fluids~\cite{YuPRE2024}. Isomorphs refer to quasi-invariant curves for certain properly reduced structural and dynamical properties in the thermodynamic phase diagram~\cite{DyreJPCB2014}. Normally, the isomorph theory applies to systems with strong correlations between the virial and potential energy equilibrium fluctuations~\cite{GnanJCP2009,DyreJPCB2014}. Such systems are called ``Roskilde-simple'' (or just ``R-simple'') and Yukawa systems have been demonstrated to be R-simple~\cite{VeldhorstPoP2015}. Among the conventional isomorph invariants are the radial distribution function, the velocity autocorrelation function, reduced transport coefficients such as self-diffusion, viscosity, and thermal conductivity, as well as some other system properties. The description of R-simple systems is greatly simplified, because their phase diagram becomes essentially one-dimensional with regard to several physical properties. Note that any isomorph invariant is in principle suitable as a system control parameter. Following Rosenfeld~\cite{RosenfeldPRA1977,RosenfeldJPCM1999}, who proposed excess entropy scaling of transport coefficients, it is conventional to use the excess entropy as a control parameter. Isomorphs are then usually considered as lines of constant excess entropy. However, any other isomorph invariant can be apparently used in place of the excess entropy, and for instance correlations between the shear viscosity and thermal conductivity coefficients have been previously reported ~\cite{KhrapakJETPLett2021}. Here we elaborate on the correlation between the thermal conductivity coefficient and the transverse sound velocity as a generalized variant of the Bridgman's formula. Note that since the conventional sound velocity $c_{\rm s}$ is {\it not an isomorph invariant}, the original Bridgman's formula does not pass an isomorph check. Not surprisingly, it is violated in systems like complex (dusty) plasma.   

The following simple qualitative argument can be put forward. Actually, heat is transferred not with the long-wavelength compressional sound velocity, but with an appropriate averaging of longitudinal and transverse phonon velocities over the effective fluid Brillouin zone. Following the standard practice in calculating the heat capacity of solids we can introduce an average sound velocity $\bar{v}$ via~\cite{LandauStatPhys}
\begin{displaymath}
\frac{3}{\bar{v}^3}=\frac{1}{c_l^3}+\frac{2}{c_t^3},
\end{displaymath}
where $c_l$ and $c_t$ are the longitudinal and transverse sound velocities, respectively. 
Since the transverse sound propagates slower and there are two transverse modes compared to one longitudinal mode, the average sound velocity should be close to the transverse one, $\bar{v}\sim c_t$. In ordinary neutral fluids $c_s$ and $c_t$ are comparable. For example, in the Lennard-Jones fluid at near freezing conditions $c_s\simeq 8 v_{\rm T}$~\cite{KhrapakPoF2023} and $c_t\simeq 6v_{\rm T}$~\cite{KhrapakMolecules2020}, where $v_{\rm T}=\sqrt{T/m}$ is the thermal velocity. In this case, one can use both $c_s$ and $c_t$ for rough estimates and Eq.~(\ref{tc1}) is appropriate. In the limit of soft interactions and plasma-related systems strong inequality $c_l\simeq c_s \gg c_t$ holds~\cite{KhrapakPoF2023} and Eq.~(\ref{tc1}) with $c_t$ instead of $c_s$ becomes more appropriate.

One more argument is based on the vibrational picture of atomic dynamics of dense liquids~\cite{KhrapakPhysRep2024}. A vibrational model of heat transfer in simple fluids with soft pairwise interactions assumes that the energy transfer rate can be approximated by the energy difference between neighbouring atoms in the direction of the temperature gradient, multiplied by the average vibrational frequency. Assuming, in addition that all atom oscillate with the same Einstein frequency, $\Omega_{\rm E}$, this leads to an expression~\cite{KhrapakPRE01_2021,Horrocks1960}
\begin{equation}
\Lambda=c_{\rm v}\frac{\Omega_{\rm E}}{2\pi \Delta}, 
\end{equation}                
where $c_{\rm v}$ is the specific heat at constant volume. This expression appears relatively accurate for strongly coupled Yukawa fluids~\cite{KhrapakPoP08_2021,KhrapakPPR2023}. In addition, in the strongly coupled limit we have $c_{\rm v}\simeq 3$ according to the Dulong-Petit law. Thus, we get $\Lambda\simeq 0.48\Omega_{\rm E}/\Delta$ in this regime. Now consider the relation between the Einstein frequency and the transverse sound velocity in Yukawa fluids. Adopting a simple stepwise approximation for the RDF~\cite{StishovJETPLett1980,GoldenPSS1993,KhrapakPoP2016}, $g(r)\simeq \theta (r-Ra)$, where $\theta(x)$ is the Heaviside step function and $R$ is the excluded hole radius expressed in units of $a$, we can get~\cite{KhrapakAIPAdv2017}       
\begin{equation}
\Omega_{\rm E}^2=\frac{\omega_{\rm p}^2}{3}e^{-R\kappa}\left(1+R\kappa\right). 
\end{equation}
Within the same approximation, the transverse sound velocity can be expressed as~\cite{KhrapakIEEE2018}
\begin{equation}
c_t^2=\frac{\omega_{\rm p}^2a^2}{30}R^2e^{-R\kappa}(1+R\kappa).
\end{equation}
Combining these expressions and taking into account that $R\simeq 1$ in the weakly screened regime 
we finally obtain
\begin{equation}
\Lambda\simeq 2.4 \frac{c_t}{\Delta^2}.
\end{equation}
Thus, a linear correlation between $\Lambda$ and $c_t$ naturally emerges within the vibrational model of atomic transport. 

\begin{figure}
\includegraphics[width=8cm]{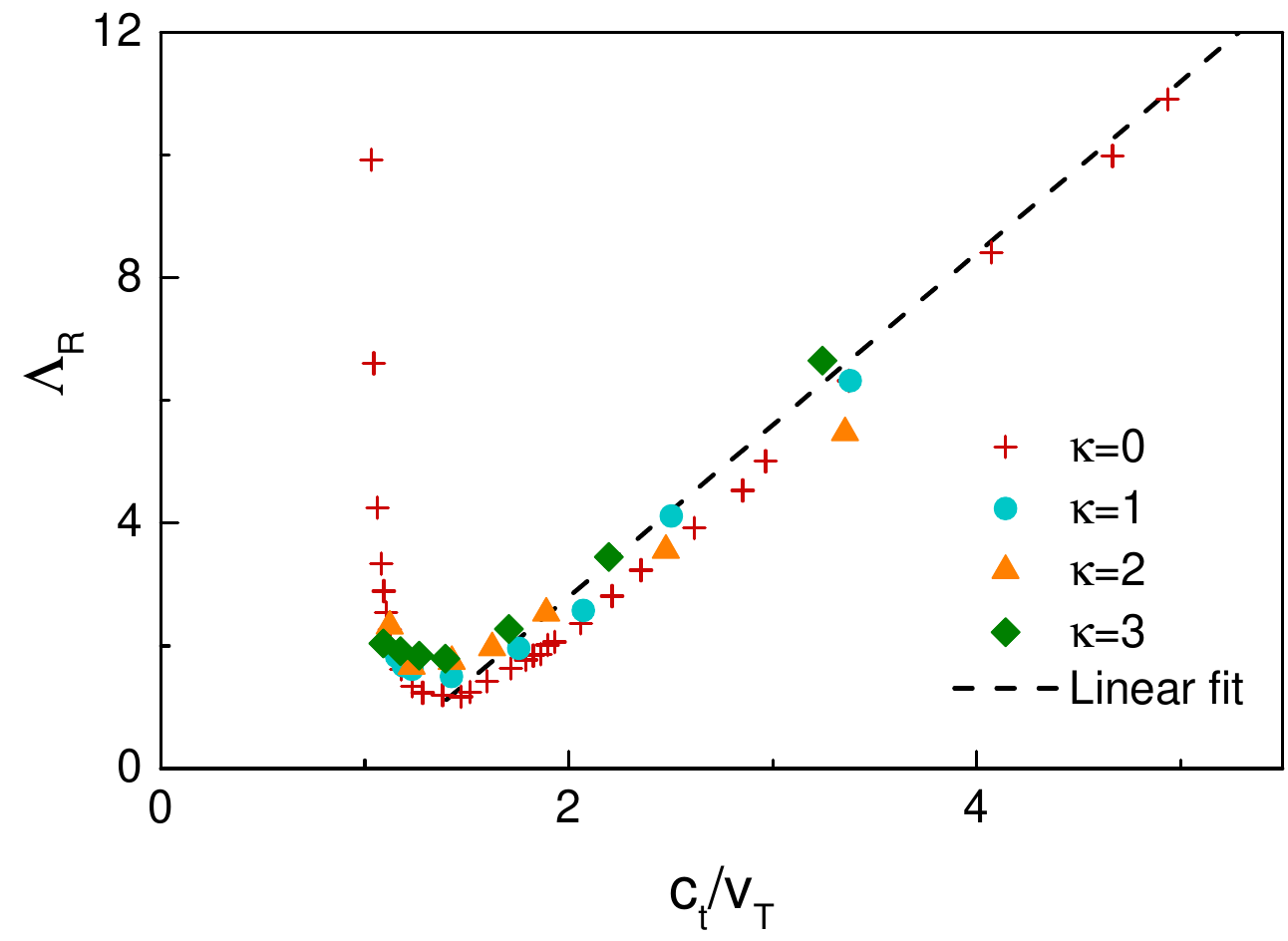}
\caption{(Color online) Reduced thermal conductivity coefficient $\Lambda_{\rm R}$  versus the reduced transverse sound velocity $c_t/v_{\rm T}$ in Yukawa fluids. The symbols correspond to numerical results from Refs.~\cite{DonkoPRE2004,ScheinerPRE2019}. The dashed curve is a linear fit of Eq.~(\ref{fit}).}
\label{lambdaR}
\end{figure}

Therefore, there are several convincing arguments to expect a correlation between the thermal conductivity coefficient and the transverse sound velocity. To verify this, it is convenient to use appropriate system-independent reduced units. For the thermal conductivity coefficient we use        
\begin{equation}\label{Rosenfeld}
\Lambda_{\rm R}=\Lambda\frac{\Delta^2}{v_{\rm T}}.
\end{equation}
This normalization is essential in the Rosenfeld's excess entropy scaling approach~\cite{RosenfeldJPCM1999}, and this is reflected by the subscript ``R''. The sound velocity is then naturally expressed in units of the thermal velocity $v_{\rm T}$. For the Yukawa fluid, the reduced transverse sound velocity is a quasi-universal function of the coupling parameter $\Gamma$, divided by its value at freezing $\Gamma_{\rm fr}$~\cite{YuPRE2024}:
\begin{equation}\label{ct}
\frac{c_t}{v_{\rm T}}\simeq \left(1+22.3\frac{\Gamma}{\Gamma_{\rm fr}}\right)^{1/2}.
\end{equation}
The values of $\Gamma_{\rm fr}$ for various values of $\kappa$ ($\leq 5$) can be found in Ref.~\cite{HamaguchiPRE1997}, or estimated from a simple analytical fit of Ref.~\cite{VaulinaJETP2000}.

The dependence of the reduced thermal conductivity coefficient of the Yukawa fluid with different $\kappa$ values on the reduced transverse sound velocity is shown in Fig.~\ref{lambdaR}. The numerical data are recalculated from those presented in Refs.~\cite{DonkoPRE2004,ScheinerPRE2019}. The following picture emerges. The thermal conductivity coefficient diverges at $c_t/v_{\rm T}=1$ and drops quickly as $c_t/v_{\rm T}$ increases. The minimum is reached at $c_t/v_{\rm T}\simeq 1.4$. Then, the thermal conductivity $\Lambda_{\rm R}$ increases monotonously with $c_t/v_{\rm T}$. At the freezing point we expect quite generally $c_t/v_{\rm T}\sim 5$~\cite{KhrapakPRR2020,YuPRE2024} and $\Lambda_{\rm R}\sim 10$ for monatomic fluids~\cite{KhrapakJMolLiq2023}. The OCP fluid data from Ref.~\cite{ScheinerPRE2019} clearly support this.        

The physics behind this behaviour is as follows. In the dilute gas regime the thermal conductivity coefficient can be estimated from the elementary kinetic formula~\cite{LifshitzKinetics} as $\Lambda\sim v_{\rm T}/\sigma$, where $\sigma$ is the effective energy transfer cross section, which is independent of the density. We get therefore $\Lambda_{\rm R}\sim 1/\sigma\rho^{2/3}$, which implies that the reduced thermal conductivity diverges as $\Lambda_{\rm R}\propto\rho^{-2/3}$ as $\rho\rightarrow 0$.   
In this regime $c_t/v_{\rm T}\simeq 1$ and we observe a rapidly decreasing asymptote to the right from $c_t/v_{\rm T}=1$. In the dense fluid regime, both $\Lambda_{\rm R}$ and $c_t/v_{\rm T}$ increase monotonously with $\Gamma$ and we observe a linear correlation between them, which can be approximately described by a simple linear fit 
\begin{equation}\label{fit}
\Lambda_{\rm R}\simeq 2.8\left( \frac{c_t}{v_{\rm T}}-1\right).
\end{equation}
The minimum in the reduced conductivity coefficient is a common property of the liquid state~\cite{KhrapakPoF2022}. It roughly corresponds to the transition between the two different mechanisms of energy transfer -- those in dilute gases and dense fluids. This transition is often referred to as the Frenkel line (FL) on the phase diagram~\cite{BrazhkinPRE2012,BrazhkinPRL2013,BrazhkinUFN2012}. Several metrics indicate that in Yukawa fluids the FL is parallel to the freezing line and is located at $\Gamma/\Gamma_{\rm fr}\simeq 0.05$~\cite{HuangPRR2023}.    
The transverse sound velocity reaches a quasi-universal value of $c_t/v_{\rm T}\simeq 1.4$ at the FL~\cite{HuangPRR2023}, which is consistent with Eq.~(\ref{ct}) above. This has been interpreted in terms of the equal contribution to the transverse sound velocity from the kinetic and potential terms at the FL~\cite{YuPRE2024}. The FL actually sets up the applicability limit of our approach: The approach is applicable at sufficiently high densities, above the density at the minimum of $\Lambda_{\rm R}$. In terms of the transverse sound velocity this corresponds to $c_t/v_{\rm T}\gtrsim 1.4$ for the Yukawa fluid. Within the applicability regime, Eq.~(\ref{fit}) is in reasonable accuracy with the available numerical data. More data points for the strongly coupled Yukawa fluid with $\kappa>0$ would be helpful to confirm this conclusion.      
      
To conclude, the conventional Bridgman's formula relating the thermal conductivity coefficient and sound velocity of dense liquids is not applicable to complex plasma fluids, because the sound velocity becomes very high in the weakly screened regime. A modification is proposed, which suggests a linear correlation between the thermal conductivity coefficient and the transverse (shear) sound velocity.
The correlation works relatively well in the strong coupling regime, above the Frenkel line. In terms of the transverse sound velocity, its applicability regime is $c_t/v_{\rm T}\gtrsim 1.4$.
An appropriate linear fit has been suggested. This correlation can be of some practical use in complex    plasmas, because the transverse sound velocity of these systems can be measured experimentally at strong coupling~\cite{PramanikPRL2002,NunomuraPRL2005,BandyopadhuayPLA2008}. The correlations between the thermal conductivity and the transverse sound velocity are likely to operate in other simple fluids, although Eq.~(\ref{fit}) might require some modification.




\bibliography{SE_Ref}

\providecommand{\noopsort}[1]{}\providecommand{\singleletter}[1]{#1}%
\begin{thebibliography}{67}%
\makeatletter
\providecommand \@ifxundefined [1]{%
 \@ifx{#1\undefined}
}%
\providecommand \@ifnum [1]{%
 \ifnum #1\expandafter \@firstoftwo
 \else \expandafter \@secondoftwo
 \fi
}%
\providecommand \@ifx [1]{%
 \ifx #1\expandafter \@firstoftwo
 \else \expandafter \@secondoftwo
 \fi
}%
\providecommand \natexlab [1]{#1}%
\providecommand \enquote  [1]{``#1''}%
\providecommand \bibnamefont  [1]{#1}%
\providecommand \bibfnamefont [1]{#1}%
\providecommand \citenamefont [1]{#1}%
\providecommand \href@noop [0]{\@secondoftwo}%
\providecommand \href [0]{\begingroup \@sanitize@url \@href}%
\providecommand \@href[1]{\@@startlink{#1}\@@href}%
\providecommand \@@href[1]{\endgroup#1\@@endlink}%
\providecommand \@sanitize@url [0]{\catcode `\\12\catcode `\$12\catcode
  `\&12\catcode `\#12\catcode `\^12\catcode `\_12\catcode `\%12\relax}%
\providecommand \@@startlink[1]{}%
\providecommand \@@endlink[0]{}%
\providecommand \url  [0]{\begingroup\@sanitize@url \@url }%
\providecommand \@url [1]{\endgroup\@href {#1}{\urlprefix }}%
\providecommand \urlprefix  [0]{URL }%
\providecommand \Eprint [0]{\href }%
\providecommand \doibase [0]{http://dx.doi.org/}%
\providecommand \selectlanguage [0]{\@gobble}%
\providecommand \bibinfo  [0]{\@secondoftwo}%
\providecommand \bibfield  [0]{\@secondoftwo}%
\providecommand \translation [1]{[#1]}%
\providecommand \BibitemOpen [0]{}%
\providecommand \bibitemStop [0]{}%
\providecommand \bibitemNoStop [0]{.\EOS\space}%
\providecommand \EOS [0]{\spacefactor3000\relax}%
\providecommand \BibitemShut  [1]{\csname bibitem#1\endcsname}%
\let\auto@bib@innerbib\@empty
\bibitem [{\citenamefont {Bridgman}(1923)}]{Bridgman1923}%
  \BibitemOpen
  \bibfield  {author} {\bibinfo {author} {\bibfnamefont {P.~W.}\ \bibnamefont
  {Bridgman}},\ }\bibfield  {title} {\enquote {\bibinfo {title} {The thermal
  conductivity of liquids under pressure},}\ }\href {\doibase 10.2307/20026073}
  {\bibfield  {journal} {\bibinfo  {journal} {PNAAS}\ }\textbf {\bibinfo
  {volume} {59}},\ \bibinfo {pages} {141} (\bibinfo {year} {1923})}\BibitemShut
  {NoStop}%
\bibitem [{\citenamefont {Bird}\ \emph {et~al.}(2002)\citenamefont {Bird},
  \citenamefont {Lightfoot},\ and\ \citenamefont {Stewart}}]{BirdBook}%
  \BibitemOpen
  \bibfield  {author} {\bibinfo {author} {\bibfnamefont {R.~B.}\ \bibnamefont
  {Bird}}, \bibinfo {author} {\bibfnamefont {E.~N.}\ \bibnamefont {Lightfoot}},
  \ and\ \bibinfo {author} {\bibfnamefont {W.~E.}\ \bibnamefont {Stewart}},\
  }\href@noop {} {\emph {\bibinfo {title} {Transport Phenomena -}}}\ (\bibinfo
  {publisher} {J. Wiley},\ \bibinfo {address} {New York},\ \bibinfo {year}
  {2002})\BibitemShut {NoStop}%
\bibitem [{\citenamefont {Xi}\ \emph {et~al.}(2020)\citenamefont {Xi},
  \citenamefont {Zhong}, \citenamefont {He}, \citenamefont {Xu}, \citenamefont
  {Nakayama}, \citenamefont {Wang}, \citenamefont {Liu}, \citenamefont {Zhou},\
  and\ \citenamefont {Li}}]{XiCPL2020}%
  \BibitemOpen
  \bibfield  {author} {\bibinfo {author} {\bibfnamefont {Q.}~\bibnamefont
  {Xi}}, \bibinfo {author} {\bibfnamefont {J.}~\bibnamefont {Zhong}}, \bibinfo
  {author} {\bibfnamefont {J.}~\bibnamefont {He}}, \bibinfo {author}
  {\bibfnamefont {X.}~\bibnamefont {Xu}}, \bibinfo {author} {\bibfnamefont
  {T.}~\bibnamefont {Nakayama}}, \bibinfo {author} {\bibfnamefont
  {Y.}~\bibnamefont {Wang}}, \bibinfo {author} {\bibfnamefont {J.}~\bibnamefont
  {Liu}}, \bibinfo {author} {\bibfnamefont {J.}~\bibnamefont {Zhou}}, \ and\
  \bibinfo {author} {\bibfnamefont {B.}~\bibnamefont {Li}},\ }\bibfield
  {title} {\enquote {\bibinfo {title} {A ubiquitous thermal conductivity
  formula for liquids, polymer glass, and amorphous solids},}\ }\href {\doibase
  10.1088/0256-307x/37/10/104401} {\bibfield  {journal} {\bibinfo  {journal}
  {Chin. Phys. Lett.}\ }\textbf {\bibinfo {volume} {37}},\ \bibinfo {pages}
  {104401} (\bibinfo {year} {2020})}\BibitemShut {NoStop}%
\bibitem [{\citenamefont {Zhao}\ \emph {et~al.}(2021)\citenamefont {Zhao},
  \citenamefont {Wingert}, \citenamefont {Chen},\ and\ \citenamefont
  {Garay}}]{ZhaoJAP2021}%
  \BibitemOpen
  \bibfield  {author} {\bibinfo {author} {\bibfnamefont {A.~Z.}\ \bibnamefont
  {Zhao}}, \bibinfo {author} {\bibfnamefont {M.~C.}\ \bibnamefont {Wingert}},
  \bibinfo {author} {\bibfnamefont {R.}~\bibnamefont {Chen}}, \ and\ \bibinfo
  {author} {\bibfnamefont {J.~E.}\ \bibnamefont {Garay}},\ }\bibfield  {title}
  {\enquote {\bibinfo {title} {Phonon gas model for thermal conductivity of
  dense, strongly interacting liquids},}\ }\href {\doibase 10.1063/5.0040734}
  {\bibfield  {journal} {\bibinfo  {journal} {J. Appl. Phys.}\ }\textbf
  {\bibinfo {volume} {129}},\ \bibinfo {pages} {235101} (\bibinfo {year}
  {2021})}\BibitemShut {NoStop}%
\bibitem [{\citenamefont {Khrapak}(2021{\natexlab{a}})}]{KhrapakPRE01_2021}%
  \BibitemOpen
  \bibfield  {author} {\bibinfo {author} {\bibfnamefont {S.~A.}\ \bibnamefont
  {Khrapak}},\ }\bibfield  {title} {\enquote {\bibinfo {title} {Vibrational
  model of thermal conduction for fluids with soft interactions},}\ }\href
  {\doibase 10.1103/physreve.103.013207} {\bibfield  {journal} {\bibinfo
  {journal} {Phys. Rev. E}\ }\textbf {\bibinfo {volume} {103}},\ \bibinfo
  {pages} {013207} (\bibinfo {year} {2021}{\natexlab{a}})}\BibitemShut
  {NoStop}%
\bibitem [{\citenamefont {Chen}(2021)}]{Chen2021}%
  \BibitemOpen
  \bibfield  {author} {\bibinfo {author} {\bibfnamefont {G.}~\bibnamefont
  {Chen}},\ }\bibfield  {title} {\enquote {\bibinfo {title} {Perspectives on
  molecular-level understanding of thermophysics of liquids and future research
  directions},}\ }\href {\doibase 10.1115/1.4052657} {\bibfield  {journal}
  {\bibinfo  {journal} {J. Heat Transf.}\ }\textbf {\bibinfo {volume} {144}},\
  \bibinfo {pages} {010801} (\bibinfo {year} {2021})}\BibitemShut {NoStop}%
\bibitem [{\citenamefont {Khrapak}(2023{\natexlab{a}})}]{KhrapakJMolLiq2023}%
  \BibitemOpen
  \bibfield  {author} {\bibinfo {author} {\bibfnamefont {S.A.}\ \bibnamefont
  {Khrapak}},\ }\bibfield  {title} {\enquote {\bibinfo {title} {Bridgman
  formula for the thermal conductivity of atomic and molecular liquids},}\
  }\href {\doibase 10.1016/j.molliq.2023.121786} {\bibfield  {journal}
  {\bibinfo  {journal} {J. Mol. Liq.}\ }\textbf {\bibinfo {volume} {381}},\
  \bibinfo {pages} {121786} (\bibinfo {year} {2023}{\natexlab{a}})}\BibitemShut
  {NoStop}%
\bibitem [{\citenamefont {{Tsytovich}}(1997)}]{TsytovichUFN1997}%
  \BibitemOpen
  \bibfield  {author} {\bibinfo {author} {\bibfnamefont {V.}~\bibnamefont
  {{Tsytovich}}},\ }\bibfield  {title} {\enquote {\bibinfo {title} {{Dust
  plasma crystals, drops, and clouds.}}}\ }\href {\doibase
  10.1070/PU1997v040n01ABEH000201} {\bibfield  {journal} {\bibinfo  {journal}
  {Phys.-Usp.}\ }\textbf {\bibinfo {volume} {40}},\ \bibinfo {pages} {53--94}
  (\bibinfo {year} {1997})}\BibitemShut {NoStop}%
\bibitem [{\citenamefont {Fortov}\ and\ \citenamefont
  {Morfill}(2019)}]{FortovBook}%
  \BibitemOpen
  \bibfield  {author} {\bibinfo {author} {\bibfnamefont {V.~E.}\ \bibnamefont
  {Fortov}}\ and\ \bibinfo {author} {\bibfnamefont {G.~E.}\ \bibnamefont
  {Morfill}},\ }\href@noop {} {\emph {\bibinfo {title} {Complex and Dusty
  Plasmas - From Laboratory to Space}}}\ (\bibinfo  {publisher} {CRC Press
  LLC},\ \bibinfo {address} {Boca Raton},\ \bibinfo {year} {2019})\BibitemShut
  {NoStop}%
\bibitem [{\citenamefont {Fortov}\ \emph {et~al.}(2004)\citenamefont {Fortov},
  \citenamefont {Khrapak}, \citenamefont {Khrapak}, \citenamefont {Molotkov},\
  and\ \citenamefont {Petrov}}]{FortovUFN}%
  \BibitemOpen
  \bibfield  {author} {\bibinfo {author} {\bibfnamefont {V.~E.}\ \bibnamefont
  {Fortov}}, \bibinfo {author} {\bibfnamefont {A.~G.}\ \bibnamefont {Khrapak}},
  \bibinfo {author} {\bibfnamefont {S.~A.}\ \bibnamefont {Khrapak}}, \bibinfo
  {author} {\bibfnamefont {V.~I.}\ \bibnamefont {Molotkov}}, \ and\ \bibinfo
  {author} {\bibfnamefont {O.~F.}\ \bibnamefont {Petrov}},\ }\bibfield  {title}
  {\enquote {\bibinfo {title} {Dusty plasmas},}\ }\href {\doibase
  10.3367/ufnr.0174.200405b.0495} {\bibfield  {journal} {\bibinfo  {journal}
  {Phys.-Usp.}\ }\textbf {\bibinfo {volume} {47}},\ \bibinfo {pages} {447 --
  492} (\bibinfo {year} {2004})}\BibitemShut {NoStop}%
\bibitem [{\citenamefont {Fortov}\ \emph {et~al.}(2005)\citenamefont {Fortov},
  \citenamefont {Ivlev}, \citenamefont {Khrapak}, \citenamefont {Khrapak},\
  and\ \citenamefont {Morfill}}]{FortovPR}%
  \BibitemOpen
  \bibfield  {author} {\bibinfo {author} {\bibfnamefont {V.~E.}\ \bibnamefont
  {Fortov}}, \bibinfo {author} {\bibfnamefont {A.~V.}\ \bibnamefont {Ivlev}},
  \bibinfo {author} {\bibfnamefont {S.~A.}\ \bibnamefont {Khrapak}}, \bibinfo
  {author} {\bibfnamefont {A.~G.}\ \bibnamefont {Khrapak}}, \ and\ \bibinfo
  {author} {\bibfnamefont {G.~E.}\ \bibnamefont {Morfill}},\ }\bibfield
  {title} {\enquote {\bibinfo {title} {Complex (dusty) plasmas: Current status,
  open issues, perspectives},}\ }\href@noop {} {\bibfield  {journal} {\bibinfo
  {journal} {Phys. Rep.}\ }\textbf {\bibinfo {volume} {421}},\ \bibinfo {pages}
  {1--103} (\bibinfo {year} {2005})}\BibitemShut {NoStop}%
\bibitem [{\citenamefont {Beckers}\ \emph {et~al.}(2023)\citenamefont
  {Beckers}, \citenamefont {Berndt}, \citenamefont {Block}, \citenamefont
  {Bonitz}, \citenamefont {Bruggeman}, \citenamefont {Couëdel}, \citenamefont
  {Delzanno}, \citenamefont {Feng}, \citenamefont {Gopalakrishnan},
  \citenamefont {Greiner}, \citenamefont {Hartmann}, \citenamefont {Horányi},
  \citenamefont {Kersten}, \citenamefont {Knapek}, \citenamefont {Konopka},
  \citenamefont {Kortshagen}, \citenamefont {Kostadinova}, \citenamefont
  {Kovačević}, \citenamefont {Krasheninnikov}, \citenamefont {Mann},
  \citenamefont {Mariotti}, \citenamefont {Matthews}, \citenamefont {Melzer},
  \citenamefont {Mikikian}, \citenamefont {Nosenko}, \citenamefont {Pustylnik},
  \citenamefont {Ratynskaia}, \citenamefont {Sankaran}, \citenamefont
  {Schneider}, \citenamefont {Thimsen}, \citenamefont {Thomas}, \citenamefont
  {Thomas}, \citenamefont {Tolias},\ and\ \citenamefont {van~de
  Kerkhof}}]{BeckersPoP2023}%
  \BibitemOpen
  \bibfield  {author} {\bibinfo {author} {\bibfnamefont {J.}~\bibnamefont
  {Beckers}}, \bibinfo {author} {\bibfnamefont {J.}~\bibnamefont {Berndt}},
  \bibinfo {author} {\bibfnamefont {D.}~\bibnamefont {Block}}, \bibinfo
  {author} {\bibfnamefont {M.}~\bibnamefont {Bonitz}}, \bibinfo {author}
  {\bibfnamefont {P.~J.}\ \bibnamefont {Bruggeman}}, \bibinfo {author}
  {\bibfnamefont {L.}~\bibnamefont {Couëdel}}, \bibinfo {author}
  {\bibfnamefont {G.~L.}\ \bibnamefont {Delzanno}}, \bibinfo {author}
  {\bibfnamefont {Y.}~\bibnamefont {Feng}}, \bibinfo {author} {\bibfnamefont
  {R.}~\bibnamefont {Gopalakrishnan}}, \bibinfo {author} {\bibfnamefont
  {F.}~\bibnamefont {Greiner}}, \bibinfo {author} {\bibfnamefont
  {P.}~\bibnamefont {Hartmann}}, \bibinfo {author} {\bibfnamefont
  {M.}~\bibnamefont {Horányi}}, \bibinfo {author} {\bibfnamefont
  {H.}~\bibnamefont {Kersten}}, \bibinfo {author} {\bibfnamefont {C.~A.}\
  \bibnamefont {Knapek}}, \bibinfo {author} {\bibfnamefont {U.}~\bibnamefont
  {Konopka}}, \bibinfo {author} {\bibfnamefont {U.}~\bibnamefont {Kortshagen}},
  \bibinfo {author} {\bibfnamefont {E.~G.}\ \bibnamefont {Kostadinova}},
  \bibinfo {author} {\bibfnamefont {E.}~\bibnamefont {Kovačević}}, \bibinfo
  {author} {\bibfnamefont {S.~I.}\ \bibnamefont {Krasheninnikov}}, \bibinfo
  {author} {\bibfnamefont {I.}~\bibnamefont {Mann}}, \bibinfo {author}
  {\bibfnamefont {D.}~\bibnamefont {Mariotti}}, \bibinfo {author}
  {\bibfnamefont {L.~S.}\ \bibnamefont {Matthews}}, \bibinfo {author}
  {\bibfnamefont {A.}~\bibnamefont {Melzer}}, \bibinfo {author} {\bibfnamefont
  {M.}~\bibnamefont {Mikikian}}, \bibinfo {author} {\bibfnamefont
  {V.}~\bibnamefont {Nosenko}}, \bibinfo {author} {\bibfnamefont {M.~Y.}\
  \bibnamefont {Pustylnik}}, \bibinfo {author} {\bibfnamefont {S.}~\bibnamefont
  {Ratynskaia}}, \bibinfo {author} {\bibfnamefont {R.~M.}\ \bibnamefont
  {Sankaran}}, \bibinfo {author} {\bibfnamefont {V.}~\bibnamefont {Schneider}},
  \bibinfo {author} {\bibfnamefont {E.~J.}\ \bibnamefont {Thimsen}}, \bibinfo
  {author} {\bibfnamefont {E.}~\bibnamefont {Thomas}}, \bibinfo {author}
  {\bibfnamefont {H.~M.}\ \bibnamefont {Thomas}}, \bibinfo {author}
  {\bibfnamefont {P.}~\bibnamefont {Tolias}}, \ and\ \bibinfo {author}
  {\bibfnamefont {M.}~\bibnamefont {van~de Kerkhof}},\ }\bibfield  {title}
  {\enquote {\bibinfo {title} {Physics and applications of dusty plasmas: The
  perspectives 2023},}\ }\href {\doibase 10.1063/5.0168088} {\bibfield
  {journal} {\bibinfo  {journal} {Phys. Plasmas}\ }\textbf {\bibinfo {volume}
  {30}},\ \bibinfo {pages} {120601} (\bibinfo {year} {2023})}\BibitemShut
  {NoStop}%
\bibitem [{\citenamefont {Nunomura}\ \emph
  {et~al.}(2005{\natexlab{a}})\citenamefont {Nunomura}, \citenamefont
  {Samsonov}, \citenamefont {Zhdanov},\ and\ \citenamefont
  {Morfill}}]{NunomuraPRL2005_HT}%
  \BibitemOpen
  \bibfield  {author} {\bibinfo {author} {\bibfnamefont {S.}~\bibnamefont
  {Nunomura}}, \bibinfo {author} {\bibfnamefont {D.}~\bibnamefont {Samsonov}},
  \bibinfo {author} {\bibfnamefont {S.}~\bibnamefont {Zhdanov}}, \ and\
  \bibinfo {author} {\bibfnamefont {G.}~\bibnamefont {Morfill}},\ }\bibfield
  {title} {\enquote {\bibinfo {title} {Heat transfer in a two-dimensional
  crystalline complex (dusty) plasma},}\ }\href {\doibase
  10.1103/physrevlett.95.025003} {\bibfield  {journal} {\bibinfo  {journal}
  {Phys. Rev. Lett.}\ }\textbf {\bibinfo {volume} {95}},\ \bibinfo {pages}
  {025003} (\bibinfo {year} {2005}{\natexlab{a}})}\BibitemShut {NoStop}%
\bibitem [{\citenamefont {Fortov}\ \emph {et~al.}(2007)\citenamefont {Fortov},
  \citenamefont {Vaulina}, \citenamefont {Petrov}, \citenamefont {Vasiliev},
  \citenamefont {Gavrikov}, \citenamefont {Shakova}, \citenamefont {Vorona},
  \citenamefont {Khrustalyov}, \citenamefont {Manohin},\ and\ \citenamefont
  {Chernyshev}}]{FortovPRE2007}%
  \BibitemOpen
  \bibfield  {author} {\bibinfo {author} {\bibfnamefont {V.~E.}\ \bibnamefont
  {Fortov}}, \bibinfo {author} {\bibfnamefont {O.~S.}\ \bibnamefont {Vaulina}},
  \bibinfo {author} {\bibfnamefont {O.~F.}\ \bibnamefont {Petrov}}, \bibinfo
  {author} {\bibfnamefont {M.~N.}\ \bibnamefont {Vasiliev}}, \bibinfo {author}
  {\bibfnamefont {A.~V.}\ \bibnamefont {Gavrikov}}, \bibinfo {author}
  {\bibfnamefont {I.~A.}\ \bibnamefont {Shakova}}, \bibinfo {author}
  {\bibfnamefont {N.~A.}\ \bibnamefont {Vorona}}, \bibinfo {author}
  {\bibfnamefont {Yu.~V.}\ \bibnamefont {Khrustalyov}}, \bibinfo {author}
  {\bibfnamefont {A.~A.}\ \bibnamefont {Manohin}}, \ and\ \bibinfo {author}
  {\bibfnamefont {A.~V.}\ \bibnamefont {Chernyshev}},\ }\bibfield  {title}
  {\enquote {\bibinfo {title} {Experimental study of the heat transport
  processes in dusty plasma fluid},}\ }\href {\doibase
  10.1103/physreve.75.026403} {\bibfield  {journal} {\bibinfo  {journal} {Phys.
  Rev. E}\ }\textbf {\bibinfo {volume} {75}},\ \bibinfo {pages} {026403}
  (\bibinfo {year} {2007})}\BibitemShut {NoStop}%
\bibitem [{\citenamefont {Nosenko}\ \emph {et~al.}(2008)\citenamefont
  {Nosenko}, \citenamefont {Zhdanov}, \citenamefont {Ivlev}, \citenamefont
  {Morfill}, \citenamefont {Goree},\ and\ \citenamefont
  {Piel}}]{NosenkoPRL2008}%
  \BibitemOpen
  \bibfield  {author} {\bibinfo {author} {\bibfnamefont {V.}~\bibnamefont
  {Nosenko}}, \bibinfo {author} {\bibfnamefont {S.}~\bibnamefont {Zhdanov}},
  \bibinfo {author} {\bibfnamefont {A.~V.}\ \bibnamefont {Ivlev}}, \bibinfo
  {author} {\bibfnamefont {G.}~\bibnamefont {Morfill}}, \bibinfo {author}
  {\bibfnamefont {J.}~\bibnamefont {Goree}}, \ and\ \bibinfo {author}
  {\bibfnamefont {A.}~\bibnamefont {Piel}},\ }\bibfield  {title} {\enquote
  {\bibinfo {title} {Heat transport in a two-dimensional complex (dusty) plasma
  at melting conditions},}\ }\href {\doibase 10.1103/physrevlett.100.025003}
  {\bibfield  {journal} {\bibinfo  {journal} {Phys. Rev. Lett.}\ }\textbf
  {\bibinfo {volume} {100}},\ \bibinfo {pages} {025003} (\bibinfo {year}
  {2008})}\BibitemShut {NoStop}%
\bibitem [{\citenamefont {Nosenko}\ \emph {et~al.}(2021)\citenamefont
  {Nosenko}, \citenamefont {Zhdanov}, \citenamefont {Pustylnik}, \citenamefont
  {Thomas}, \citenamefont {Lipaev},\ and\ \citenamefont
  {Novitskii}}]{NosenkoPoP2021}%
  \BibitemOpen
  \bibfield  {author} {\bibinfo {author} {\bibfnamefont {V.}~\bibnamefont
  {Nosenko}}, \bibinfo {author} {\bibfnamefont {S.}~\bibnamefont {Zhdanov}},
  \bibinfo {author} {\bibfnamefont {M.}~\bibnamefont {Pustylnik}}, \bibinfo
  {author} {\bibfnamefont {H.~M.}\ \bibnamefont {Thomas}}, \bibinfo {author}
  {\bibfnamefont {A.~M.}\ \bibnamefont {Lipaev}}, \ and\ \bibinfo {author}
  {\bibfnamefont {O.~V.}\ \bibnamefont {Novitskii}},\ }\bibfield  {title}
  {\enquote {\bibinfo {title} {Heat transport in a flowing complex plasma in
  microgravity conditions},}\ }\href {\doibase 10.1063/5.0069672} {\bibfield
  {journal} {\bibinfo  {journal} {Phys. Plasmas}\ }\textbf {\bibinfo {volume}
  {28}},\ \bibinfo {pages} {113701} (\bibinfo {year} {2021})}\BibitemShut
  {NoStop}%
\bibitem [{\citenamefont {Brush}\ \emph {et~al.}(1966)\citenamefont {Brush},
  \citenamefont {Sahlin},\ and\ \citenamefont {Teller}}]{BrushJCP1966}%
  \BibitemOpen
  \bibfield  {author} {\bibinfo {author} {\bibfnamefont {S.~G.}\ \bibnamefont
  {Brush}}, \bibinfo {author} {\bibfnamefont {H.~L.}\ \bibnamefont {Sahlin}}, \
  and\ \bibinfo {author} {\bibfnamefont {E.}~\bibnamefont {Teller}},\
  }\bibfield  {title} {\enquote {\bibinfo {title} {Monte {C}arlo study of a
  one-component plasma},}\ }\href {\doibase 10.1063/1.1727895} {\bibfield
  {journal} {\bibinfo  {journal} {J. Chem. Phys.}\ }\textbf {\bibinfo {volume}
  {45}},\ \bibinfo {pages} {2102--2118} (\bibinfo {year} {1966})}\BibitemShut
  {NoStop}%
\bibitem [{\citenamefont {Baus}\ and\ \citenamefont
  {Hansen}(1980)}]{BausPR1980}%
  \BibitemOpen
  \bibfield  {author} {\bibinfo {author} {\bibfnamefont {M}~\bibnamefont
  {Baus}}\ and\ \bibinfo {author} {\bibfnamefont {J.~P.}\ \bibnamefont
  {Hansen}},\ }\bibfield  {title} {\enquote {\bibinfo {title} {Statistical
  mechanics of simple {C}oulomb systems},}\ }\href {\doibase
  10.1016/0370-1573(80)90022-8} {\bibfield  {journal} {\bibinfo  {journal}
  {Phys. Rep.}\ }\textbf {\bibinfo {volume} {59}},\ \bibinfo {pages} {1--94}
  (\bibinfo {year} {1980})}\BibitemShut {NoStop}%
\bibitem [{\citenamefont {Khrapak}\ and\ \citenamefont
  {Thomas}(2015)}]{KhrapakPRE03_2015}%
  \BibitemOpen
  \bibfield  {author} {\bibinfo {author} {\bibfnamefont {S.~A.}\ \bibnamefont
  {Khrapak}}\ and\ \bibinfo {author} {\bibfnamefont {H.~M.}\ \bibnamefont
  {Thomas}},\ }\bibfield  {title} {\enquote {\bibinfo {title} {Fluid approach
  to evaluate sound velocity in {Y}ukawa systems and complex plasmas},}\ }\href
  {\doibase 10.1103/physreve.91.033110} {\bibfield  {journal} {\bibinfo
  {journal} {Phys. Rev. E}\ }\textbf {\bibinfo {volume} {91}},\ \bibinfo
  {pages} {033110} (\bibinfo {year} {2015})}\BibitemShut {NoStop}%
\bibitem [{\citenamefont {Khrapak}(2015)}]{KhrapakPPCF2015}%
  \BibitemOpen
  \bibfield  {author} {\bibinfo {author} {\bibfnamefont {S.~A.}\ \bibnamefont
  {Khrapak}},\ }\bibfield  {title} {\enquote {\bibinfo {title} {Thermodynamics
  of {Y}ukawa systems and sound velocity in dusty plasmas},}\ }\href {\doibase
  10.1088/0741-3335/58/1/014022} {\bibfield  {journal} {\bibinfo  {journal}
  {Plasma Phys. Control. Fusion}\ }\textbf {\bibinfo {volume} {58}},\ \bibinfo
  {pages} {014022} (\bibinfo {year} {2015})}\BibitemShut {NoStop}%
\bibitem [{\citenamefont {Khrapak}(2019)}]{KhrapakPoP10_2019}%
  \BibitemOpen
  \bibfield  {author} {\bibinfo {author} {\bibfnamefont {S.~A.}\ \bibnamefont
  {Khrapak}},\ }\bibfield  {title} {\enquote {\bibinfo {title} {Unified
  description of sound velocities in strongly coupled {Y}ukawa systems of
  different spatial dimensionality},}\ }\href {\doibase 10.1063/1.5124676}
  {\bibfield  {journal} {\bibinfo  {journal} {Physics of Plasmas}\ }\textbf
  {\bibinfo {volume} {26}},\ \bibinfo {pages} {103703} (\bibinfo {year}
  {2019})}\BibitemShut {NoStop}%
\bibitem [{\citenamefont {Rao}\ \emph {et~al.}(1990)\citenamefont {Rao},
  \citenamefont {Shukla},\ and\ \citenamefont {Yu}}]{Rao1990}%
  \BibitemOpen
  \bibfield  {author} {\bibinfo {author} {\bibfnamefont {N.N.}\ \bibnamefont
  {Rao}}, \bibinfo {author} {\bibfnamefont {P.K.}\ \bibnamefont {Shukla}}, \
  and\ \bibinfo {author} {\bibfnamefont {M.Y.}\ \bibnamefont {Yu}},\ }\bibfield
   {title} {\enquote {\bibinfo {title} {Dust-acoustic waves in dusty
  plasmas},}\ }\href {\doibase 10.1016/0032-0633(90)90147-i} {\bibfield
  {journal} {\bibinfo  {journal} {Planet. Space Sci.}\ }\textbf {\bibinfo
  {volume} {38}},\ \bibinfo {pages} {543} (\bibinfo {year} {1990})}\BibitemShut
  {NoStop}%
\bibitem [{\citenamefont {Ohta}\ and\ \citenamefont
  {Hamaguchi}(2000)}]{OhtaPRL2000}%
  \BibitemOpen
  \bibfield  {author} {\bibinfo {author} {\bibfnamefont {H.}~\bibnamefont
  {Ohta}}\ and\ \bibinfo {author} {\bibfnamefont {S.}~\bibnamefont
  {Hamaguchi}},\ }\bibfield  {title} {\enquote {\bibinfo {title} {Wave
  dispersion relations in {Y}ukawa fluids},}\ }\href {\doibase
  10.1103/physrevlett.84.6026} {\bibfield  {journal} {\bibinfo  {journal}
  {Phys. Rev. Lett.}\ }\textbf {\bibinfo {volume} {84}},\ \bibinfo {pages}
  {6026--6029} (\bibinfo {year} {2000})}\BibitemShut {NoStop}%
\bibitem [{\citenamefont {Scheiner}\ and\ \citenamefont
  {Baalrud}(2019)}]{ScheinerPRE2019}%
  \BibitemOpen
  \bibfield  {author} {\bibinfo {author} {\bibfnamefont {B.}~\bibnamefont
  {Scheiner}}\ and\ \bibinfo {author} {\bibfnamefont {S.~D.}\ \bibnamefont
  {Baalrud}},\ }\bibfield  {title} {\enquote {\bibinfo {title} {Testing thermal
  conductivity models with equilibrium molecular dynamics simulations of the
  one-component plasma},}\ }\href {\doibase 10.1103/physreve.100.043206}
  {\bibfield  {journal} {\bibinfo  {journal} {Phys. Rev. E}\ }\textbf {\bibinfo
  {volume} {100}},\ \bibinfo {pages} {043206} (\bibinfo {year}
  {2019})}\BibitemShut {NoStop}%
\bibitem [{\citenamefont {Pramanik}\ \emph {et~al.}(2002)\citenamefont
  {Pramanik}, \citenamefont {Prasad}, \citenamefont {Sen},\ and\ \citenamefont
  {Kaw}}]{PramanikPRL2002}%
  \BibitemOpen
  \bibfield  {author} {\bibinfo {author} {\bibfnamefont {J.}~\bibnamefont
  {Pramanik}}, \bibinfo {author} {\bibfnamefont {G.}~\bibnamefont {Prasad}},
  \bibinfo {author} {\bibfnamefont {A.}~\bibnamefont {Sen}}, \ and\ \bibinfo
  {author} {\bibfnamefont {P.~K.}\ \bibnamefont {Kaw}},\ }\bibfield  {title}
  {\enquote {\bibinfo {title} {Experimental observations of transverse shear
  waves in strongly coupled dusty plasmas},}\ }\href {\doibase
  10.1103/physrevlett.88.175001} {\bibfield  {journal} {\bibinfo  {journal}
  {Phys. Rev. Lett.}\ }\textbf {\bibinfo {volume} {88}},\ \bibinfo {pages}
  {175001} (\bibinfo {year} {2002})}\BibitemShut {NoStop}%
\bibitem [{\citenamefont {Nunomura}\ \emph
  {et~al.}(2005{\natexlab{b}})\citenamefont {Nunomura}, \citenamefont
  {Zhdanov}, \citenamefont {Samsonov},\ and\ \citenamefont
  {Morfill}}]{NunomuraPRL2005}%
  \BibitemOpen
  \bibfield  {author} {\bibinfo {author} {\bibfnamefont {S.}~\bibnamefont
  {Nunomura}}, \bibinfo {author} {\bibfnamefont {S.}~\bibnamefont {Zhdanov}},
  \bibinfo {author} {\bibfnamefont {D.}~\bibnamefont {Samsonov}}, \ and\
  \bibinfo {author} {\bibfnamefont {G.}~\bibnamefont {Morfill}},\ }\bibfield
  {title} {\enquote {\bibinfo {title} {Wave spectra in solid and liquid complex
  (dusty) plasmas},}\ }\href {\doibase 10.1103/physrevlett.94.045001}
  {\bibfield  {journal} {\bibinfo  {journal} {Phys. Rev. Lett.}\ }\textbf
  {\bibinfo {volume} {94}},\ \bibinfo {pages} {045001} (\bibinfo {year}
  {2005}{\natexlab{b}})}\BibitemShut {NoStop}%
\bibitem [{\citenamefont {Bandyopadhyay}\ \emph {et~al.}(2008)\citenamefont
  {Bandyopadhyay}, \citenamefont {Prasad}, \citenamefont {Sen},\ and\
  \citenamefont {Kaw}}]{BandyopadhuayPLA2008}%
  \BibitemOpen
  \bibfield  {author} {\bibinfo {author} {\bibfnamefont {P.}~\bibnamefont
  {Bandyopadhyay}}, \bibinfo {author} {\bibfnamefont {G.}~\bibnamefont
  {Prasad}}, \bibinfo {author} {\bibfnamefont {A.}~\bibnamefont {Sen}}, \ and\
  \bibinfo {author} {\bibfnamefont {P.K.}\ \bibnamefont {Kaw}},\ }\bibfield
  {title} {\enquote {\bibinfo {title} {Driven transverse shear waves in a
  strongly coupled dusty plasma},}\ }\href {\doibase
  10.1016/j.physleta.2008.06.051} {\bibfield  {journal} {\bibinfo  {journal}
  {Phys. Lett. A}\ }\textbf {\bibinfo {volume} {372}},\ \bibinfo {pages} {5467}
  (\bibinfo {year} {2008})}\BibitemShut {NoStop}%
\bibitem [{\citenamefont {Hosokawa}\ \emph {et~al.}(2009)\citenamefont
  {Hosokawa}, \citenamefont {Inui}, \citenamefont {Kajihara}, \citenamefont
  {Matsuda}, \citenamefont {Ichitsubo}, \citenamefont {Pilgrim}, \citenamefont
  {Sinn}, \citenamefont {Gonz{\'{a}}lez}, \citenamefont {Gonz{\'{a}}lez},
  \citenamefont {Tsutsui},\ and\ \citenamefont {Baron}}]{HosokawaPRL2009}%
  \BibitemOpen
  \bibfield  {author} {\bibinfo {author} {\bibfnamefont {S.}~\bibnamefont
  {Hosokawa}}, \bibinfo {author} {\bibfnamefont {M.}~\bibnamefont {Inui}},
  \bibinfo {author} {\bibfnamefont {Y.}~\bibnamefont {Kajihara}}, \bibinfo
  {author} {\bibfnamefont {K.}~\bibnamefont {Matsuda}}, \bibinfo {author}
  {\bibfnamefont {T.}~\bibnamefont {Ichitsubo}}, \bibinfo {author}
  {\bibfnamefont {W.-C.}\ \bibnamefont {Pilgrim}}, \bibinfo {author}
  {\bibfnamefont {H.}~\bibnamefont {Sinn}}, \bibinfo {author} {\bibfnamefont
  {L.~E.}\ \bibnamefont {Gonz{\'{a}}lez}}, \bibinfo {author} {\bibfnamefont
  {D.~J.}\ \bibnamefont {Gonz{\'{a}}lez}}, \bibinfo {author} {\bibfnamefont
  {S.}~\bibnamefont {Tsutsui}}, \ and\ \bibinfo {author} {\bibfnamefont
  {A.~Q.~R.}\ \bibnamefont {Baron}},\ }\bibfield  {title} {\enquote {\bibinfo
  {title} {Transverse acoustic excitations in liquid {G}a},}\ }\href {\doibase
  10.1103/physrevlett.102.105502} {\bibfield  {journal} {\bibinfo  {journal}
  {Phys. Rev. Lett.}\ }\textbf {\bibinfo {volume} {102}},\ \bibinfo {pages}
  {105502} (\bibinfo {year} {2009})}\BibitemShut {NoStop}%
\bibitem [{\citenamefont {Hosokawa}\ \emph {et~al.}(2015)\citenamefont
  {Hosokawa}, \citenamefont {Inui}, \citenamefont {Kajihara}, \citenamefont
  {Tsutsui},\ and\ \citenamefont {Baron}}]{HosokawaJPCM2015}%
  \BibitemOpen
  \bibfield  {author} {\bibinfo {author} {\bibfnamefont {S.}~\bibnamefont
  {Hosokawa}}, \bibinfo {author} {\bibfnamefont {M.}~\bibnamefont {Inui}},
  \bibinfo {author} {\bibfnamefont {Y.}~\bibnamefont {Kajihara}}, \bibinfo
  {author} {\bibfnamefont {S.}~\bibnamefont {Tsutsui}}, \ and\ \bibinfo
  {author} {\bibfnamefont {A.~Q.~R.}\ \bibnamefont {Baron}},\ }\bibfield
  {title} {\enquote {\bibinfo {title} {Transverse excitations in liquid {Fe},
  {Cu} and {Zn}},}\ }\href {\doibase 10.1088/0953-8984/27/19/194104} {\bibfield
   {journal} {\bibinfo  {journal} {J. Phys.: Condens. Matter}\ }\textbf
  {\bibinfo {volume} {27}},\ \bibinfo {pages} {194104} (\bibinfo {year}
  {2015})}\BibitemShut {NoStop}%
\bibitem [{\citenamefont {Goree}\ \emph {et~al.}(2012)\citenamefont {Goree},
  \citenamefont {Donk{\'{o}}},\ and\ \citenamefont {Hartmann}}]{GoreePRE2012}%
  \BibitemOpen
  \bibfield  {author} {\bibinfo {author} {\bibfnamefont {J.}~\bibnamefont
  {Goree}}, \bibinfo {author} {\bibfnamefont {Z.}~\bibnamefont {Donk{\'{o}}}},
  \ and\ \bibinfo {author} {\bibfnamefont {P.}~\bibnamefont {Hartmann}},\
  }\bibfield  {title} {\enquote {\bibinfo {title} {Cutoff wave number for shear
  waves and {M}axwell relaxation time in {Y}ukawa liquids},}\ }\href {\doibase
  10.1103/physreve.85.066401} {\bibfield  {journal} {\bibinfo  {journal} {Phys.
  Rev. E}\ }\textbf {\bibinfo {volume} {85}},\ \bibinfo {pages} {066401}
  (\bibinfo {year} {2012})}\BibitemShut {NoStop}%
\bibitem [{\citenamefont {Bolmatov}\ \emph {et~al.}(2016)\citenamefont
  {Bolmatov}, \citenamefont {Zhernenkov}, \citenamefont {Zav’yalov},
  \citenamefont {Stoupin}, \citenamefont {Cunsolo},\ and\ \citenamefont
  {Cai}}]{BolmatovSciRep2016}%
  \BibitemOpen
  \bibfield  {author} {\bibinfo {author} {\bibfnamefont {D.}~\bibnamefont
  {Bolmatov}}, \bibinfo {author} {\bibfnamefont {M.}~\bibnamefont
  {Zhernenkov}}, \bibinfo {author} {\bibfnamefont {D.}~\bibnamefont
  {Zav’yalov}}, \bibinfo {author} {\bibfnamefont {S.}~\bibnamefont
  {Stoupin}}, \bibinfo {author} {\bibfnamefont {A.}~\bibnamefont {Cunsolo}}, \
  and\ \bibinfo {author} {\bibfnamefont {Y.~Q.}\ \bibnamefont {Cai}},\
  }\bibfield  {title} {\enquote {\bibinfo {title} {Thermally triggered phononic
  gaps in liquids at {TH}z scale},}\ }\href {\doibase 10.1038/srep19469}
  {\bibfield  {journal} {\bibinfo  {journal} {Sci. Rep.}\ }\textbf {\bibinfo
  {volume} {6}},\ \bibinfo {pages} {19469} (\bibinfo {year}
  {2016})}\BibitemShut {NoStop}%
\bibitem [{\citenamefont {Bryk}\ \emph {et~al.}(2017)\citenamefont {Bryk},
  \citenamefont {Huerta}, \citenamefont {Hordiichuk},\ and\ \citenamefont
  {Trokhymchuk}}]{BrykJCP2017}%
  \BibitemOpen
  \bibfield  {author} {\bibinfo {author} {\bibfnamefont {T.}~\bibnamefont
  {Bryk}}, \bibinfo {author} {\bibfnamefont {A.}~\bibnamefont {Huerta}},
  \bibinfo {author} {\bibfnamefont {V.}~\bibnamefont {Hordiichuk}}, \ and\
  \bibinfo {author} {\bibfnamefont {A.~D.}\ \bibnamefont {Trokhymchuk}},\
  }\bibfield  {title} {\enquote {\bibinfo {title} {Non-hydrodynamic transverse
  collective excitations in hard-sphere fluids},}\ }\href {\doibase
  10.1063/1.4997640} {\bibfield  {journal} {\bibinfo  {journal} {J. Chem.
  Phys.}\ }\textbf {\bibinfo {volume} {147}},\ \bibinfo {pages} {064509}
  (\bibinfo {year} {2017})}\BibitemShut {NoStop}%
\bibitem [{\citenamefont {Yang}\ \emph {et~al.}(2017)\citenamefont {Yang},
  \citenamefont {Dove}, \citenamefont {Brazhkin},\ and\ \citenamefont
  {Trachenko}}]{YangPRL2017}%
  \BibitemOpen
  \bibfield  {author} {\bibinfo {author} {\bibfnamefont {C.}~\bibnamefont
  {Yang}}, \bibinfo {author} {\bibfnamefont {M.~T.}\ \bibnamefont {Dove}},
  \bibinfo {author} {\bibfnamefont {V.~V.}\ \bibnamefont {Brazhkin}}, \ and\
  \bibinfo {author} {\bibfnamefont {K.}~\bibnamefont {Trachenko}},\ }\bibfield
  {title} {\enquote {\bibinfo {title} {Emergence and evolution of the k-gap in
  spectra of liquid and supercritical states},}\ }\href {\doibase
  10.1103/physrevlett.118.215502} {\bibfield  {journal} {\bibinfo  {journal}
  {Phys. Rev. Lett.}\ }\textbf {\bibinfo {volume} {118}},\ \bibinfo {pages}
  {215502} (\bibinfo {year} {2017})}\BibitemShut {NoStop}%
\bibitem [{\citenamefont {Kryuchkov}\ \emph {et~al.}(2019)\citenamefont
  {Kryuchkov}, \citenamefont {Mistryukova}, \citenamefont {Brazhkin},\ and\
  \citenamefont {Yurchenko}}]{KryuchkovSciRep2019}%
  \BibitemOpen
  \bibfield  {author} {\bibinfo {author} {\bibfnamefont {N.~P.}\ \bibnamefont
  {Kryuchkov}}, \bibinfo {author} {\bibfnamefont {L.~A.}\ \bibnamefont
  {Mistryukova}}, \bibinfo {author} {\bibfnamefont {V.~V.}\ \bibnamefont
  {Brazhkin}}, \ and\ \bibinfo {author} {\bibfnamefont {S.~O.}\ \bibnamefont
  {Yurchenko}},\ }\bibfield  {title} {\enquote {\bibinfo {title} {Excitation
  spectra in fluids: How to analyze them properly},}\ }\href {\doibase
  10.1038/s41598-019-46979-y} {\bibfield  {journal} {\bibinfo  {journal} {Sci.
  Rep.}\ }\textbf {\bibinfo {volume} {9}},\ \bibinfo {pages} {10483} (\bibinfo
  {year} {2019})}\BibitemShut {NoStop}%
\bibitem [{\citenamefont {Khrapak}\ \emph {et~al.}(2019)\citenamefont
  {Khrapak}, \citenamefont {Khrapak}, \citenamefont {Kryuchkov},\ and\
  \citenamefont {Yurchenko}}]{KhrapakJCP2019}%
  \BibitemOpen
  \bibfield  {author} {\bibinfo {author} {\bibfnamefont {S.~A.}\ \bibnamefont
  {Khrapak}}, \bibinfo {author} {\bibfnamefont {A.~G.}\ \bibnamefont
  {Khrapak}}, \bibinfo {author} {\bibfnamefont {N.~P.}\ \bibnamefont
  {Kryuchkov}}, \ and\ \bibinfo {author} {\bibfnamefont {S.~O.}\ \bibnamefont
  {Yurchenko}},\ }\bibfield  {title} {\enquote {\bibinfo {title} {Onset of
  transverse (shear) waves in strongly-coupled {Y}ukawa fluids},}\ }\href
  {\doibase 10.1063/1.5088141} {\bibfield  {journal} {\bibinfo  {journal} {J.
  Chem. Phys.}\ }\textbf {\bibinfo {volume} {150}},\ \bibinfo {pages} {104503}
  (\bibinfo {year} {2019})}\BibitemShut {NoStop}%
\bibitem [{\citenamefont {Zwanzig}\ and\ \citenamefont
  {Mountain}(1965)}]{ZwanzigJCP1965}%
  \BibitemOpen
  \bibfield  {author} {\bibinfo {author} {\bibfnamefont {R.}~\bibnamefont
  {Zwanzig}}\ and\ \bibinfo {author} {\bibfnamefont {R.~D.}\ \bibnamefont
  {Mountain}},\ }\bibfield  {title} {\enquote {\bibinfo {title} {High-frequency
  elastic moduli of simple fluids},}\ }\href {\doibase 10.1063/1.1696718}
  {\bibfield  {journal} {\bibinfo  {journal} {J. Chem. Phys.}\ }\textbf
  {\bibinfo {volume} {43}},\ \bibinfo {pages} {4464--4471} (\bibinfo {year}
  {1965})}\BibitemShut {NoStop}%
\bibitem [{\citenamefont {Khrapak}\ and\ \citenamefont
  {Khrapak}(2018)}]{KhrapakIEEE2018}%
  \BibitemOpen
  \bibfield  {author} {\bibinfo {author} {\bibfnamefont {S.}~\bibnamefont
  {Khrapak}}\ and\ \bibinfo {author} {\bibfnamefont {A.}~\bibnamefont
  {Khrapak}},\ }\bibfield  {title} {\enquote {\bibinfo {title} {Simple
  dispersion relations for {C}oulomb and {Y}ukawa fluids},}\ }\href {\doibase
  10.1109/tps.2017.2763741} {\bibfield  {journal} {\bibinfo  {journal} {{IEEE}
  Trans. Plasma Sci.}\ }\textbf {\bibinfo {volume} {46}},\ \bibinfo {pages}
  {737--742} (\bibinfo {year} {2018})}\BibitemShut {NoStop}%
\bibitem [{\citenamefont {Khrapak}\ and\ \citenamefont
  {Klumov}(2020)}]{KhrapakPoP02_2020}%
  \BibitemOpen
  \bibfield  {author} {\bibinfo {author} {\bibfnamefont {S.~A.}\ \bibnamefont
  {Khrapak}}\ and\ \bibinfo {author} {\bibfnamefont {B.~A.}\ \bibnamefont
  {Klumov}},\ }\bibfield  {title} {\enquote {\bibinfo {title} {Instantaneous
  shear modulus of {Y}ukawa fluids across coupling regimes},}\ }\href {\doibase
  10.1063/1.5140858} {\bibfield  {journal} {\bibinfo  {journal} {Physics of
  Plasmas}\ }\textbf {\bibinfo {volume} {27}},\ \bibinfo {pages} {024501}
  (\bibinfo {year} {2020})}\BibitemShut {NoStop}%
\bibitem [{\citenamefont {Khrapak}(2020{\natexlab{a}})}]{KhrapakPRR2020}%
  \BibitemOpen
  \bibfield  {author} {\bibinfo {author} {\bibfnamefont {S.~A.}\ \bibnamefont
  {Khrapak}},\ }\bibfield  {title} {\enquote {\bibinfo {title} {Lindemann
  melting criterion in two dimensions},}\ }\href {\doibase
  10.1103/physrevresearch.2.012040} {\bibfield  {journal} {\bibinfo  {journal}
  {Phys. Rev. Research}\ }\textbf {\bibinfo {volume} {2}},\ \bibinfo {pages}
  {012040} (\bibinfo {year} {2020}{\natexlab{a}})}\BibitemShut {NoStop}%
\bibitem [{\citenamefont {Khrapak}\ and\ \citenamefont
  {Khrapak}(2022{\natexlab{a}})}]{KhrapakPoF2022}%
  \BibitemOpen
  \bibfield  {author} {\bibinfo {author} {\bibfnamefont {S.~A.}\ \bibnamefont
  {Khrapak}}\ and\ \bibinfo {author} {\bibfnamefont {A.~G.}\ \bibnamefont
  {Khrapak}},\ }\bibfield  {title} {\enquote {\bibinfo {title} {Minima of shear
  viscosity and thermal conductivity coefficients of classical fluids},}\
  }\href {\doibase 10.1063/5.0082465} {\bibfield  {journal} {\bibinfo
  {journal} {Phys. Fluids}\ }\textbf {\bibinfo {volume} {34}},\ \bibinfo
  {pages} {027102} (\bibinfo {year} {2022}{\natexlab{a}})}\BibitemShut
  {NoStop}%
\bibitem [{\citenamefont {Khrapak}\ and\ \citenamefont
  {Khrapak}(2022{\natexlab{b}})}]{KhrapakJCP2022_1}%
  \BibitemOpen
  \bibfield  {author} {\bibinfo {author} {\bibfnamefont {S.~A.}\ \bibnamefont
  {Khrapak}}\ and\ \bibinfo {author} {\bibfnamefont {A.~G.}\ \bibnamefont
  {Khrapak}},\ }\bibfield  {title} {\enquote {\bibinfo {title} {Freezing
  density scaling of fluid transport properties: Application to liquefied noble
  gases},}\ }\href {\doibase 10.1063/5.0096947} {\bibfield  {journal} {\bibinfo
   {journal} {J. Chem. Phys.}\ }\textbf {\bibinfo {volume} {157}},\ \bibinfo
  {pages} {014501} (\bibinfo {year} {2022}{\natexlab{b}})}\BibitemShut
  {NoStop}%
\bibitem [{\citenamefont {Yu}\ \emph {et~al.}(2024)\citenamefont {Yu},
  \citenamefont {Huang}, \citenamefont {Lu}, \citenamefont {Khrapak},\ and\
  \citenamefont {Feng}}]{YuPRE2024}%
  \BibitemOpen
  \bibfield  {author} {\bibinfo {author} {\bibfnamefont {N.}~\bibnamefont
  {Yu}}, \bibinfo {author} {\bibfnamefont {D.}~\bibnamefont {Huang}}, \bibinfo
  {author} {\bibfnamefont {S.}~\bibnamefont {Lu}}, \bibinfo {author}
  {\bibfnamefont {S.}~\bibnamefont {Khrapak}}, \ and\ \bibinfo {author}
  {\bibfnamefont {Y.}~\bibnamefont {Feng}},\ }\bibfield  {title} {\enquote
  {\bibinfo {title} {Universal scaling of transverse sound speed and its
  isomorphic property in {Y}ukawa fluids},}\ }\href {\doibase
  10.1103/physreve.109.035202} {\bibfield  {journal} {\bibinfo  {journal}
  {Phys. Rev. E}\ }\textbf {\bibinfo {volume} {109}},\ \bibinfo {pages}
  {035202} (\bibinfo {year} {2024})}\BibitemShut {NoStop}%
\bibitem [{\citenamefont {Dyre}(2014)}]{DyreJPCB2014}%
  \BibitemOpen
  \bibfield  {author} {\bibinfo {author} {\bibfnamefont {J.~C.}\ \bibnamefont
  {Dyre}},\ }\bibfield  {title} {\enquote {\bibinfo {title} {Hidden scale
  invariance in condensed matter},}\ }\href {\doibase 10.1021/jp501852b}
  {\bibfield  {journal} {\bibinfo  {journal} {J. Phys. Chem. B}\ }\textbf
  {\bibinfo {volume} {118}},\ \bibinfo {pages} {10007--10024} (\bibinfo {year}
  {2014})}\BibitemShut {NoStop}%
\bibitem [{\citenamefont {Gnan}\ \emph {et~al.}(2009)\citenamefont {Gnan},
  \citenamefont {Schroder}, \citenamefont {Pedersen}, \citenamefont {Bailey},\
  and\ \citenamefont {Dyre}}]{GnanJCP2009}%
  \BibitemOpen
  \bibfield  {author} {\bibinfo {author} {\bibfnamefont {N.}~\bibnamefont
  {Gnan}}, \bibinfo {author} {\bibfnamefont {T.~B.}\ \bibnamefont {Schroder}},
  \bibinfo {author} {\bibfnamefont {U.~R.}\ \bibnamefont {Pedersen}}, \bibinfo
  {author} {\bibfnamefont {N.~P.}\ \bibnamefont {Bailey}}, \ and\ \bibinfo
  {author} {\bibfnamefont {J.~C.}\ \bibnamefont {Dyre}},\ }\bibfield  {title}
  {\enquote {\bibinfo {title} {Pressure-energy correlations in liquids. {IV}.
  {I}somorphs in liquid phase diagrams},}\ }\href {\doibase 10.1063/1.3265957}
  {\bibfield  {journal} {\bibinfo  {journal} {J. Chem. Phys.}\ }\textbf
  {\bibinfo {volume} {131}},\ \bibinfo {pages} {234504} (\bibinfo {year}
  {2009})}\BibitemShut {NoStop}%
\bibitem [{\citenamefont {Veldhorst}\ \emph {et~al.}(2015)\citenamefont
  {Veldhorst}, \citenamefont {Schroder},\ and\ \citenamefont
  {Dyre}}]{VeldhorstPoP2015}%
  \BibitemOpen
  \bibfield  {author} {\bibinfo {author} {\bibfnamefont {A.~A.}\ \bibnamefont
  {Veldhorst}}, \bibinfo {author} {\bibfnamefont {T.~B.}\ \bibnamefont
  {Schroder}}, \ and\ \bibinfo {author} {\bibfnamefont {J.~C.}\ \bibnamefont
  {Dyre}},\ }\bibfield  {title} {\enquote {\bibinfo {title} {Invariants in the
  {Y}ukawa system thermodynamic phase diagram},}\ }\href {\doibase
  10.1063/1.4926822} {\bibfield  {journal} {\bibinfo  {journal} {Phys.
  Plasmas}\ }\textbf {\bibinfo {volume} {22}},\ \bibinfo {pages} {073705}
  (\bibinfo {year} {2015})}\BibitemShut {NoStop}%
\bibitem [{\citenamefont {Rosenfeld}(1977)}]{RosenfeldPRA1977}%
  \BibitemOpen
  \bibfield  {author} {\bibinfo {author} {\bibfnamefont {Y.}~\bibnamefont
  {Rosenfeld}},\ }\bibfield  {title} {\enquote {\bibinfo {title} {Relation
  between the transport coefficients and the internal entropy of simple
  systems},}\ }\href {\doibase 10.1103/physreva.15.2545} {\bibfield  {journal}
  {\bibinfo  {journal} {Phys. Rev. A}\ }\textbf {\bibinfo {volume} {15}},\
  \bibinfo {pages} {2545--2549} (\bibinfo {year} {1977})}\BibitemShut {NoStop}%
\bibitem [{\citenamefont {Rosenfeld}(1999)}]{RosenfeldJPCM1999}%
  \BibitemOpen
  \bibfield  {author} {\bibinfo {author} {\bibfnamefont {Y.}~\bibnamefont
  {Rosenfeld}},\ }\bibfield  {title} {\enquote {\bibinfo {title} {A
  quasi-universal scaling law for atomic transport in simple fluids},}\ }\href
  {\doibase 10.1088/0953-8984/11/28/303} {\bibfield  {journal} {\bibinfo
  {journal} {J. Phys.: Condens. Matter}\ }\textbf {\bibinfo {volume} {11}},\
  \bibinfo {pages} {5415--5427} (\bibinfo {year} {1999})}\BibitemShut {NoStop}%
\bibitem [{\citenamefont {Khrapak}\ and\ \citenamefont
  {Khrapak}(2021)}]{KhrapakJETPLett2021}%
  \BibitemOpen
  \bibfield  {author} {\bibinfo {author} {\bibfnamefont {S.~A.}\ \bibnamefont
  {Khrapak}}\ and\ \bibinfo {author} {\bibfnamefont {A.~G.}\ \bibnamefont
  {Khrapak}},\ }\bibfield  {title} {\enquote {\bibinfo {title} {Correlations
  between the shear viscosity and thermal conductivity coefficients of dense
  simple liquids},}\ }\href {\doibase 10.1134/s0021364021210037} {\bibfield
  {journal} {\bibinfo  {journal} {{JETP} Lett.}\ }\textbf {\bibinfo {volume}
  {114}},\ \bibinfo {pages} {540} (\bibinfo {year} {2021})}\BibitemShut
  {NoStop}%
\bibitem [{\citenamefont {Landau}\ \emph {et~al.}(1980)\citenamefont {Landau},
  \citenamefont {Lifshic},\ and\ \citenamefont {Pitaevskii}}]{LandauStatPhys}%
  \BibitemOpen
  \bibfield  {author} {\bibinfo {author} {\bibfnamefont {L.~D.}\ \bibnamefont
  {Landau}}, \bibinfo {author} {\bibfnamefont {E.~M.}\ \bibnamefont {Lifshic}},
  \ and\ \bibinfo {author} {\bibfnamefont {L.~P.}\ \bibnamefont {Pitaevskii}},\
  }\href@noop {} {\emph {\bibinfo {title} {Statistical Physics}}}\ (\bibinfo
  {publisher} {Butterworth-Heinemann},\ \bibinfo {address} {Oxford},\ \bibinfo
  {year} {1980})\BibitemShut {NoStop}%
\bibitem [{\citenamefont {Khrapak}\ and\ \citenamefont
  {Khrapak}(2023)}]{KhrapakPoF2023}%
  \BibitemOpen
  \bibfield  {author} {\bibinfo {author} {\bibfnamefont {S.~A.}\ \bibnamefont
  {Khrapak}}\ and\ \bibinfo {author} {\bibfnamefont {A.~G.}\ \bibnamefont
  {Khrapak}},\ }\bibfield  {title} {\enquote {\bibinfo {title} {Sound
  velocities in liquids near freezing: {D}ependence on the interaction
  potential and correlations with thermal conductivity},}\ }\href {\doibase
  10.1063/5.0157945} {\bibfield  {journal} {\bibinfo  {journal} {Phys. Fluids}\
  }\textbf {\bibinfo {volume} {35}},\ \bibinfo {pages} {077129} (\bibinfo
  {year} {2023})}\BibitemShut {NoStop}%
\bibitem [{\citenamefont {Khrapak}(2020{\natexlab{b}})}]{KhrapakMolecules2020}%
  \BibitemOpen
  \bibfield  {author} {\bibinfo {author} {\bibfnamefont {S.~A.}\ \bibnamefont
  {Khrapak}},\ }\bibfield  {title} {\enquote {\bibinfo {title} {Sound
  velocities of {L}ennard-{J}ones systems near the liquid-solid phase
  transition},}\ }\href {\doibase 10.3390/molecules25153498} {\bibfield
  {journal} {\bibinfo  {journal} {Molecules}\ }\textbf {\bibinfo {volume}
  {25}},\ \bibinfo {pages} {3498} (\bibinfo {year}
  {2020}{\natexlab{b}})}\BibitemShut {NoStop}%
\bibitem [{\citenamefont {Khrapak}(2024)}]{KhrapakPhysRep2024}%
  \BibitemOpen
  \bibfield  {author} {\bibinfo {author} {\bibfnamefont {S.A.}\ \bibnamefont
  {Khrapak}},\ }\bibfield  {title} {\enquote {\bibinfo {title} {Elementary
  vibrational model for transport properties of dense fluids},}\ }\href
  {\doibase 10.1016/j.physrep.2023.11.004} {\bibfield  {journal} {\bibinfo
  {journal} {Phys. Rep.}\ }\textbf {\bibinfo {volume} {1050}},\ \bibinfo
  {pages} {1} (\bibinfo {year} {2024})}\BibitemShut {NoStop}%
\bibitem [{\citenamefont {Horrocks}\ and\ \citenamefont
  {McLaughlin}(1960)}]{Horrocks1960}%
  \BibitemOpen
  \bibfield  {author} {\bibinfo {author} {\bibfnamefont {J.~K.}\ \bibnamefont
  {Horrocks}}\ and\ \bibinfo {author} {\bibfnamefont {E.}~\bibnamefont
  {McLaughlin}},\ }\bibfield  {title} {\enquote {\bibinfo {title} {Thermal
  conductivity of simple molecules in the condensed state},}\ }\href {\doibase
  10.1039/tf9605600206} {\bibfield  {journal} {\bibinfo  {journal} {Trans.
  Faraday Soc.}\ }\textbf {\bibinfo {volume} {56}},\ \bibinfo {pages} {206}
  (\bibinfo {year} {1960})}\BibitemShut {NoStop}%
\bibitem [{\citenamefont {Khrapak}(2021{\natexlab{b}})}]{KhrapakPoP08_2021}%
  \BibitemOpen
  \bibfield  {author} {\bibinfo {author} {\bibfnamefont {S.~A.}\ \bibnamefont
  {Khrapak}},\ }\bibfield  {title} {\enquote {\bibinfo {title} {Thermal
  conductivity of strongly coupled {Y}ukawa fluids},}\ }\href {\doibase
  10.1063/5.0056763} {\bibfield  {journal} {\bibinfo  {journal} {Phys.
  Plasmas}\ }\textbf {\bibinfo {volume} {28}},\ \bibinfo {pages} {084501}
  (\bibinfo {year} {2021}{\natexlab{b}})}\BibitemShut {NoStop}%
\bibitem [{\citenamefont {Khrapak}(2023{\natexlab{b}})}]{KhrapakPPR2023}%
  \BibitemOpen
  \bibfield  {author} {\bibinfo {author} {\bibfnamefont {S.~A.}\ \bibnamefont
  {Khrapak}},\ }\bibfield  {title} {\enquote {\bibinfo {title} {Vibrational
  model of heat transfer in strongly coupled {Y}ukawa fluids (dusty plasma
  liquids)},}\ }\href {\doibase 10.1134/s1063780x22600876} {\bibfield
  {journal} {\bibinfo  {journal} {Plasma Phys. Rep.}\ }\textbf {\bibinfo
  {volume} {49}},\ \bibinfo {pages} {15--22} (\bibinfo {year}
  {2023}{\natexlab{b}})}\BibitemShut {NoStop}%
\bibitem [{\citenamefont {Stishov}(1980)}]{StishovJETPLett1980}%
  \BibitemOpen
  \bibfield  {author} {\bibinfo {author} {\bibfnamefont {S.~M.}\ \bibnamefont
  {Stishov}},\ }\bibfield  {title} {\enquote {\bibinfo {title} {On the question
  of the specific heat of liquids and dense gases at low temperatures},}\
  }\href {http://jetpletters.ru/ps/0/article_20291.shtml} {\bibfield  {journal}
  {\bibinfo  {journal} {JETP Lett.}\ }\textbf {\bibinfo {volume} {31}},\
  \bibinfo {pages} {249} (\bibinfo {year} {1980})}\BibitemShut {NoStop}%
\bibitem [{\citenamefont {Golden}\ and\ \citenamefont
  {Kalman}(1993)}]{GoldenPSS1993}%
  \BibitemOpen
  \bibfield  {author} {\bibinfo {author} {\bibfnamefont {K.~I.}\ \bibnamefont
  {Golden}}\ and\ \bibinfo {author} {\bibfnamefont {G.}~\bibnamefont
  {Kalman}},\ }\bibfield  {title} {\enquote {\bibinfo {title} {Correlations
  destroy acoustic plasmons in superlattices},}\ }\href {\doibase
  10.1002/pssb.2221800224} {\bibfield  {journal} {\bibinfo  {journal} {Phys.
  Status Solidi (b)}\ }\textbf {\bibinfo {volume} {180}},\ \bibinfo {pages}
  {533} (\bibinfo {year} {1993})}\BibitemShut {NoStop}%
\bibitem [{\citenamefont {Khrapak}\ \emph {et~al.}(2016)\citenamefont
  {Khrapak}, \citenamefont {Klumov}, \citenamefont {Couedel},\ and\
  \citenamefont {Thomas}}]{KhrapakPoP2016}%
  \BibitemOpen
  \bibfield  {author} {\bibinfo {author} {\bibfnamefont {S.~A.}\ \bibnamefont
  {Khrapak}}, \bibinfo {author} {\bibfnamefont {B.}~\bibnamefont {Klumov}},
  \bibinfo {author} {\bibfnamefont {L.}~\bibnamefont {Couedel}}, \ and\
  \bibinfo {author} {\bibfnamefont {H.~M.}\ \bibnamefont {Thomas}},\ }\bibfield
   {title} {\enquote {\bibinfo {title} {On the long-waves dispersion in
  {Y}ukawa systems},}\ }\href {\doibase 10.1063/1.4942169} {\bibfield
  {journal} {\bibinfo  {journal} {Phys. Plasmas}\ }\textbf {\bibinfo {volume}
  {23}},\ \bibinfo {pages} {023702} (\bibinfo {year} {2016})}\BibitemShut
  {NoStop}%
\bibitem [{\citenamefont {Khrapak}(2017)}]{KhrapakAIPAdv2017}%
  \BibitemOpen
  \bibfield  {author} {\bibinfo {author} {\bibfnamefont {S.~A.}\ \bibnamefont
  {Khrapak}},\ }\bibfield  {title} {\enquote {\bibinfo {title} {Practical
  dispersion relations for strongly coupled plasma fluids},}\ }\href {\doibase
  10.1063/1.5002130} {\bibfield  {journal} {\bibinfo  {journal} {{AIP} Adv.}\
  }\textbf {\bibinfo {volume} {7}},\ \bibinfo {pages} {125026} (\bibinfo {year}
  {2017})}\BibitemShut {NoStop}%
\bibitem [{\citenamefont {Donk{\'{o}}}\ and\ \citenamefont
  {Hartmann}(2004)}]{DonkoPRE2004}%
  \BibitemOpen
  \bibfield  {author} {\bibinfo {author} {\bibfnamefont {Z.}~\bibnamefont
  {Donk{\'{o}}}}\ and\ \bibinfo {author} {\bibfnamefont {P.}~\bibnamefont
  {Hartmann}},\ }\bibfield  {title} {\enquote {\bibinfo {title} {Thermal
  conductivity of strongly coupled {Y}ukawa liquids},}\ }\href {\doibase
  10.1103/physreve.69.016405} {\bibfield  {journal} {\bibinfo  {journal} {Phys.
  Rev. E}\ }\textbf {\bibinfo {volume} {69}},\ \bibinfo {pages} {016405}
  (\bibinfo {year} {2004})}\BibitemShut {NoStop}%
\bibitem [{\citenamefont {Hamaguchi}\ \emph {et~al.}(1997)\citenamefont
  {Hamaguchi}, \citenamefont {Farouki},\ and\ \citenamefont
  {Dubin}}]{HamaguchiPRE1997}%
  \BibitemOpen
  \bibfield  {author} {\bibinfo {author} {\bibfnamefont {S.}~\bibnamefont
  {Hamaguchi}}, \bibinfo {author} {\bibfnamefont {R.~T.}\ \bibnamefont
  {Farouki}}, \ and\ \bibinfo {author} {\bibfnamefont {D.~H.~E.}\ \bibnamefont
  {Dubin}},\ }\bibfield  {title} {\enquote {\bibinfo {title} {Triple point of
  {Y}ukawa systems},}\ }\href {\doibase 10.1103/physreve.56.4671} {\bibfield
  {journal} {\bibinfo  {journal} {Phys. Rev. E}\ }\textbf {\bibinfo {volume}
  {56}},\ \bibinfo {pages} {4671--4682} (\bibinfo {year} {1997})}\BibitemShut
  {NoStop}%
\bibitem [{\citenamefont {Vaulina}\ and\ \citenamefont
  {Khrapak}(2000)}]{VaulinaJETP2000}%
  \BibitemOpen
  \bibfield  {author} {\bibinfo {author} {\bibfnamefont {O.~S.}\ \bibnamefont
  {Vaulina}}\ and\ \bibinfo {author} {\bibfnamefont {S.~A.}\ \bibnamefont
  {Khrapak}},\ }\bibfield  {title} {\enquote {\bibinfo {title} {Scaling law for
  the fluid-solid phase transition in {Y}ukawa systems (dusty plasmas)},}\
  }\href {\doibase 10.1134/1.559102} {\bibfield  {journal} {\bibinfo  {journal}
  {JETP}\ }\textbf {\bibinfo {volume} {90}},\ \bibinfo {pages} {287--289}
  (\bibinfo {year} {2000})}\BibitemShut {NoStop}%
\bibitem [{\citenamefont {Lifshitz}\ and\ \citenamefont
  {Pitaevskii}(1995)}]{LifshitzKinetics}%
  \BibitemOpen
  \bibfield  {author} {\bibinfo {author} {\bibfnamefont {E.M.}\ \bibnamefont
  {Lifshitz}}\ and\ \bibinfo {author} {\bibfnamefont {L.~P.}\ \bibnamefont
  {Pitaevskii}},\ }\href@noop {} {\emph {\bibinfo {title} {Physical
  Kinetics}}}\ (\bibinfo  {publisher} {Elsevier Science},\ \bibinfo {address}
  {Stanford},\ \bibinfo {year} {1995})\BibitemShut {NoStop}%
\bibitem [{\citenamefont {Brazhkin}\ \emph
  {et~al.}(2012{\natexlab{a}})\citenamefont {Brazhkin}, \citenamefont {Fomin},
  \citenamefont {Lyapin}, \citenamefont {Ryzhov},\ and\ \citenamefont
  {Trachenko}}]{BrazhkinPRE2012}%
  \BibitemOpen
  \bibfield  {author} {\bibinfo {author} {\bibfnamefont {V.~V.}\ \bibnamefont
  {Brazhkin}}, \bibinfo {author} {\bibfnamefont {Yu.~D.}\ \bibnamefont
  {Fomin}}, \bibinfo {author} {\bibfnamefont {A.~G.}\ \bibnamefont {Lyapin}},
  \bibinfo {author} {\bibfnamefont {V.~N.}\ \bibnamefont {Ryzhov}}, \ and\
  \bibinfo {author} {\bibfnamefont {K.}~\bibnamefont {Trachenko}},\ }\bibfield
  {title} {\enquote {\bibinfo {title} {Two liquid states of matter: A dynamic
  line on a phase diagram},}\ }\href {\doibase 10.1103/physreve.85.031203}
  {\bibfield  {journal} {\bibinfo  {journal} {Phys. Rev. E}\ }\textbf {\bibinfo
  {volume} {85}},\ \bibinfo {pages} {031203} (\bibinfo {year}
  {2012}{\natexlab{a}})}\BibitemShut {NoStop}%
\bibitem [{\citenamefont {Brazhkin}\ \emph {et~al.}(2013)\citenamefont
  {Brazhkin}, \citenamefont {Fomin}, \citenamefont {Lyapin}, \citenamefont
  {Ryzhov}, \citenamefont {Tsiok},\ and\ \citenamefont
  {Trachenko}}]{BrazhkinPRL2013}%
  \BibitemOpen
  \bibfield  {author} {\bibinfo {author} {\bibfnamefont {V.~V.}\ \bibnamefont
  {Brazhkin}}, \bibinfo {author} {\bibfnamefont {Yu.~D.}\ \bibnamefont
  {Fomin}}, \bibinfo {author} {\bibfnamefont {A.~G.}\ \bibnamefont {Lyapin}},
  \bibinfo {author} {\bibfnamefont {V.~N.}\ \bibnamefont {Ryzhov}}, \bibinfo
  {author} {\bibfnamefont {E.~N.}\ \bibnamefont {Tsiok}}, \ and\ \bibinfo
  {author} {\bibfnamefont {K.}~\bibnamefont {Trachenko}},\ }\bibfield  {title}
  {\enquote {\bibinfo {title} {Liquid-gas'' transition in the supercritical
  region: fundamental changes in the particle dynamics},}\ }\href@noop {}
  {\bibfield  {journal} {\bibinfo  {journal} {Phys. Rev. Lett.}\ }\textbf
  {\bibinfo {volume} {111}},\ \bibinfo {pages} {145901} (\bibinfo {year}
  {2013})}\BibitemShut {NoStop}%
\bibitem [{\citenamefont {Brazhkin}\ \emph
  {et~al.}(2012{\natexlab{b}})\citenamefont {Brazhkin}, \citenamefont {Lyapin},
  \citenamefont {Ryzhov}, \citenamefont {Trachenko}, \citenamefont {Fomin},\
  and\ \citenamefont {Tsiok}}]{BrazhkinUFN2012}%
  \BibitemOpen
  \bibfield  {author} {\bibinfo {author} {\bibfnamefont {V.~V.}\ \bibnamefont
  {Brazhkin}}, \bibinfo {author} {\bibfnamefont {A.G.}\ \bibnamefont {Lyapin}},
  \bibinfo {author} {\bibfnamefont {V.~N.}\ \bibnamefont {Ryzhov}}, \bibinfo
  {author} {\bibfnamefont {K.}~\bibnamefont {Trachenko}}, \bibinfo {author}
  {\bibfnamefont {Y.~D.}\ \bibnamefont {Fomin}}, \ and\ \bibinfo {author}
  {\bibfnamefont {E.~N.}\ \bibnamefont {Tsiok}},\ }\bibfield  {title} {\enquote
  {\bibinfo {title} {Where is the supercritical fluid on the phase diagram?}}\
  }\href {\doibase 10.3367/ufnr.0182.201211a.1137} {\bibfield  {journal}
  {\bibinfo  {journal} {Phys.-Usp.}\ }\textbf {\bibinfo {volume} {182}},\
  \bibinfo {pages} {1137--1156} (\bibinfo {year}
  {2012}{\natexlab{b}})}\BibitemShut {NoStop}%
\bibitem [{\citenamefont {Huang}\ \emph {et~al.}(2023)\citenamefont {Huang},
  \citenamefont {Baggioli}, \citenamefont {Lu}, \citenamefont {Ma},\ and\
  \citenamefont {Feng}}]{HuangPRR2023}%
  \BibitemOpen
  \bibfield  {author} {\bibinfo {author} {\bibfnamefont {D.}~\bibnamefont
  {Huang}}, \bibinfo {author} {\bibfnamefont {M.}~\bibnamefont {Baggioli}},
  \bibinfo {author} {\bibfnamefont {S.}~\bibnamefont {Lu}}, \bibinfo {author}
  {\bibfnamefont {Z.}~\bibnamefont {Ma}}, \ and\ \bibinfo {author}
  {\bibfnamefont {Y.}~\bibnamefont {Feng}},\ }\bibfield  {title} {\enquote
  {\bibinfo {title} {Revealing the supercritical dynamics of dusty plasmas and
  their liquidlike to gaslike dynamical crossover},}\ }\href {\doibase
  10.1103/physrevresearch.5.013149} {\bibfield  {journal} {\bibinfo  {journal}
  {Phys. Rev. Research}\ }\textbf {\bibinfo {volume} {5}},\ \bibinfo {pages}
  {013149} (\bibinfo {year} {2023})}\BibitemShut {NoStop}%
\end{thebibliography}%

\end{document}